\documentclass[prd, twocolumn,floats,floatfix,nofootinbib,10pt] {revtex4-1}
\usepackage{graphicx}
\usepackage{dcolumn}
\usepackage{bm}
\usepackage{graphics}
\usepackage{slashed}
\usepackage{amssymb}
\usepackage{natbib}
\usepackage{amsmath}
\usepackage{url}
\usepackage[usenames,dvipsnames,svgnames,table]{xcolor}
\newcommand{\bea}{\begin{eqnarray}}
\newcommand{\ena}{\end{eqnarray}}
\newcommand{\bean}{\begin{eqnarray*}}
\newcommand{\enan}{\end{eqnarray*}}

\begin{document}

\title{Nonlinear electrodynamics is skilled with knots}

\author{E. Goulart}
\affiliation{\textit{Centro Brasileiro de Pesquisas F\'isicas - CBPF\\Rua Dr. Xavier Sigaud, 150, CEP 22290-180, Rio de Janeiro, Brazil}\\
\texttt{egoulart@cbpf.br}}

\maketitle

\textbf{The aims of this letter are three-fold: First is to show that nonlinear generalizations of electrodynamics support various types of knotted solutions in vacuum. The solutions are universal in the sense that they do not depend on the specific Lagrangian density, at least if the latter gives rise to a well-posed theory. Second is to describe the interaction between probe waves and knotted background configurations. We show that the qualitative behaviour of this interaction may be described in terms of Robinson congruences, which appear explicitly in the causal structure of the theory. Finally, we argue that optical arrangements endowed with intense background fields could be the natural place to look for the knots experimentally.}\\

\textbf{Introduction} Knot theory is a fascinating branch of topology with many potential applications to physics \cite{Ada}. Roughly speaking, a knot is a closed loop of string, with no free ends. Mathematically, we can think in it as an embedding of a circle in 3-dimensional Euclidean space $\mathbb{R}^{3}$ which cannot be untied without cutting or permitting the curve to pass through itself. Two of such embeddings are said to represent the same knot if there exists a deformation of $\mathbb{R}^{3}$ upon itself taking one of the knots into the other smoothly (ambient isotopy). A link by its turn is a collection of knotted loops which do not intersect, all tangled up together. 

Thought-provoking is the existence of space-filling knots in some field theories with physical interest, both quantum and classical. Recently, these exotic configurations have been investigated in contexts as diverse as hydrodynamics \cite{Klech}, Bose-Einstein condensates \cite{Hall}, ferromagnetism \cite{Coo} and non-abelian gauge theories, where they are supposed to describe stable excitations such as glueballs \cite{Fad}. Little-known, however, is the unexpected result of Ra\~nada that Maxwell's linear equations admit solutions such that all the lines of force are closed and any pair of magnetic (electric) lines are linked \cite{Ran1,Ran2,Ran3,Ran4}. Ra\~nada's method is based on maps from Minkowski spacetime to the complex plane $(\varphi,\theta):\mathbb{R}^{1+3}\rightarrow \mathbb{C}$ which are required to be single-valued at spatial infinity. After identifying the complex plane with the unit sphere via stereographic projection, he effectively obtains $(\varphi,\theta): \mathbb{R}\times\mathbb{S}^{3}\rightarrow\mathbb{S}^{2}$ which are nontrivial, since the third homotopy group of the sphere is $\pi_{3}(\mathbb{S}^{2})\in\mathbb{Z}$. The linked lines are interpreted as the pre-images of the maps and fulllfill the whole of space with a fibered structure remarkably similar to the Hopf fibration \cite{Hop}. The corresponding solutions have been dubbed \textit{electromagnetic knots} and received considerable attention in the last few years. Importantly, it was shown by Irvine and Bowmeester that approximate knots of light could be generated using tightly focused circularly polarized laser beams \cite{Irv2}.


This paper deals with electromagnetic knots in the context of nonlinear electrodynamics \cite{NED,NED1}. Specifically, we show that, if the theory arises from a Lagrangian and the latter gives rise to well-behaved equations, knots and links appear as universal exact solutions.  This means that closed knotted/linked loops made of electric and magnetic fields are consistent with the nonlinear evolution equations whilst their topology are preserved in time, very much in the same way as they do in Maxwell's theory. As is well known, nonlinear theories of this type are relevant for several reasons. In QED, the polarization of the vacuum leads naturally to nonlinear effects (such as the light-light scattering) which are effectively described by Euler-Heisenberg's Lagrangian \cite{Dunne}. In dielectrics, the interaction between molecules and external fields can be described by an effective nonlinear theory, which is typically observed at very high light intensities such as those provided by pulsed lasers \cite{Yuen}. Born-Infeld model has mathematical connections to string theory, for its Lagrangian appears in relation with the gauge fields on a D-brane \cite{GG}. Finally, the increasing availability of multi-hundred TW and PW lasers brings the confirmation of long-predicted nonlinear phenomena closer \cite{laser1}, \cite{laser2}, \cite{laser3}.

Our construction is based on \textit{null fields} and employs some of the machinery recently discussed in \cite{Ioa, Ked, Hoy}. However, due to the nonlinearity of the underlying equations, there is a striking novelty in our approach: small amplitude/high frequency waves arriving from the vacuum may interact with the knotted fields in a nontrivial way. This result promises to open up the possibility of detection using ordinary radiation, which will receive an imprint of knottedness due to nonlinear effects if the background fields are sufficiently intense. \\

\textbf{Nonlinear electrodynamics} Consider a  Minkowski spacetime $(\textbf{M}, g)$ and write $F_{ab}$ for the electromagnetic field strength. Let
\begin{equation}
F:=F_{ab}F^{ab}\quad\quad G:={}^{\star}F_{ab}F^{ab}
\end{equation}
denote the field invariants, with ${}^{\star}F_{ab}$ the dual tensor. We are interested in nonlinear theories in empty space provided by the action
\begin{equation}
S=\int\mathcal{L}(F,G)\sqrt{-g}\ d^{4}x,
\end{equation}
with the Lagrangian density an arbitrary smooth function of the invariants. The equations of motion read as
\begin{equation}\label{NLE}
\nabla_{a}\big(\mathcal{L}_{F}F^{ab}+\mathcal{L}_{G}{}^{\star}F^{ab}\big)=0,\quad \nabla_{[a}F_{bc]}=0,
\end{equation}
with $\mathcal{L}_{F}$ and $\mathcal{L}_{G}$ the partial derivatives w.r.t. the invariants for conciseness. As is well known, Eqs. (4) constitute a system of first-order quasilinear PDE's with constraints and can be considerably difficult to solve in general situations. For the resulting theory to have predictive power, however, we assume henceforth: i) Well-posedness of the Cauchy problem\footnote{Well-posedness is at the roots of physics, for it amounts to the predictability power of the theory, asserting that solutions exist, are unique and depend continuously on the data (see, for instance, \cite{Reula}).}; ii) Maxwell's electrodynamics is recovered for sufficiently weak fields. 

In order to talk about electric and magnetic fields a spacetime foliation by the level sets $\Sigma_{t}$ of a scalar $t(x)$ is required. Here the gradient $t_{a}:=\partial_{a}t$ is taken to be timelike, future-directed and normalized everywhere. With these conventions we may uniquely decompose the field strength as
\begin{equation}
F_{ab}=g_{ab}^{\phantom a\phantom a cd}E_{c}t_{d}+\eta_{ab}^{\phantom a\phantom a cd}B_{c}t_{d},
\end{equation}
with $g_{abcd}:=g_{ac}g_{bd}-g_{ad}g_{bc}$, $\eta_{abcd}$ the completely antisymmetric Levi-Civita tensor and
\begin{equation}
E_{a}=F_{a}^{\phantom a b}t_{b},\quad\quad B_{a}=-{}^{\star}F_{a}^{\phantom a b}t_{b}
\end{equation}
the electromagnetic fields. The concept of space-filling lines of force thus appears as the integral curves in $\Sigma_{t}$ satisfying the system
\begin{equation}\nonumber
\frac{dx^{a}}{d\lambda}=E^{a}(x(\lambda)),\quad\quad \frac{dx^{a}}{d\tau}=B^{a}(x(\tau)),
\end{equation}
with $\tau$ and $\lambda$ real parameters. Knowledge of the geometry/topology of these lines for all $t$ is sufficient to qualitatively describe the fields satisfying (\ref{NLE}) in time. This framework naturally induces the question:  Are there solutions of Eqs. (\ref{NLE}) such that the corresponding field lines form knotted loops filling the hyper-surfaces $\Sigma_{t}$ with a nontrivial topology? Surprisingly, we shall see that such solutions exist, at least for fields satisfying the \textit{null condition} (FIG. 1).\\
 
\begin{figure}[h]
       \centering  
       \includegraphics[scale=0.47]{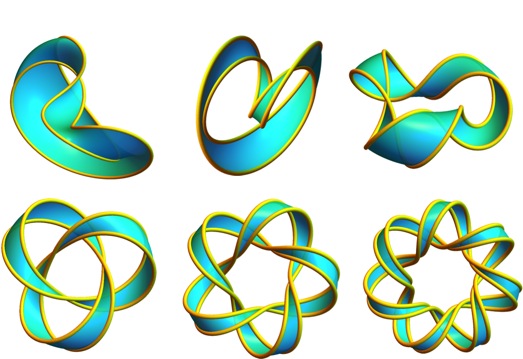}
       \caption{Examples of topologically nontrivial field lines (depicted in yellow) at a given time. For the sake of illustration, we sketched also the Seifert surfaces (depicted in blue) joining the field lines. The three configurations above are linked but not knotted whereas the three below are linked and knotted.}
       \label{dynamic}
\end{figure} 
 
\textbf{From complex mappings to null fields} A powerful (though not very well known) mechanism to generate null electromagnetic fields dates back to Bateman \cite{Bat}. This approach has been recently employed by Besieris and Shaarawi \cite{Ioa}, Kedia \textit{et al} \cite{Ked} and Hoyos \textit{et al} \cite{Hoy}. Let us, very briefly, review some aspects of the formalism and describe it in a more covariant fashion. Consider, to begin with, complex maps $\alpha,\beta: (\textbf{M},g)\rightarrow\mathbb{C}$ satisfying the fully nonlinear system of first order PDE's
\begin{equation}\label{Bateman1}
\big(g^{abcd}+i\eta^{abcd}\big)\partial_{c}\alpha\partial_{d}\beta=0.
\end{equation}
It is known that a sufficient condition for (\ref{Bateman1}) to admit nontrivial solutions is that
\begin{equation}\label{Bateman2}
(\partial_{a}\alpha\partial^{a}\alpha)(\partial_{b}\beta\partial^{b}\beta)=0,\quad\quad \partial_{a}\alpha\partial^{a}\beta=0.
\end{equation}
A pair $\alpha,\beta$ solving (\ref{Bateman1}) and (\ref{Bateman2}) is called a \textit{Bateman pair} and induces the antisymmetric complex tensor
\begin{equation}\label{Silberstein}
\mathcal{M}_{ab}:=\eta_{ab}^{\phantom a\phantom a cd}\partial_{c}\alpha\partial_{d}\beta,
\end{equation}
called the \textit{Riemann-Silberstein} 2-form. With this definition one can easily verify that the imaginary part of the Riemann-Silberstein 2-form is minus the dual of its real part, i.e.
\begin{equation}\label{EM}
\mathcal{M}^{ab}=F^{ab}-i{}^{\star}F^{ab}.
\end{equation}
A field $F_{ab}$ specified in this way is necessarily null (self-conjugate in Bateman's terminology), for if we contract (\ref{EM}) with itself we obtain
\begin{equation}
\mathcal{M}^{ab}\mathcal{M}_{ab}=2(F-iG)=0,
\end{equation}
which identically vanishes as a consequence of (\ref{Bateman1}). 

A remarkable property of this construction is that it automatically generates null solutions to Maxwell's linear equations in vacuum. Indeed, applying the covariant divergence to (\ref{EM}) and using (\ref{Bateman1}), there follows
\begin{eqnarray*}
\nabla_{a}\mathcal{M}^{ab}&=&\nabla_{a}F^{ab}-i\nabla_{a}{}^{\star}F^{ab}=0.
\end{eqnarray*}
From this analysis it is clear that if a pair $\alpha,\beta$ is known, it is always possible to obtain algorithmically a null solution of Maxwell's equations in vacuum. What is more, once the pair is obtained, a whole family of new pairs $f$, $g$ result, where $f$ and $g$ can be taken as arbitrary holomorphic functions of $\alpha$ and $\beta$ (see reference \cite{Ked} for further details).

What about the equations of nonlinear electrodynamics? Can we extend the technique discussed so far so as to find solutions of (\ref{NLE})? The answer is provided by the following lemma: \textit{Every Bateman pair automatically induces null solutions of the nonlinear equations} (\ref{NLE}). The proof is straightforward since if the field is null everywhere, the quantities $\mathcal{L}_{F}$ and $\mathcal{L}_{G}$ in Eqs. (\ref{NLE}) can be treated as constants and, therefore, drop out from the covariant derivatives, implying again:
\begin{eqnarray}
\nabla_{a}F^{ab}=0,\quad\quad \nabla_{a}{}^{\star}F^{ab}=0,
\end{eqnarray}
which are nothing but Maxwell's equations in free-space.\\

\textbf{Knotted and linked solutions} An interesting consequence of the lemma is that a class of well-known knotted solutions discussed in the literature may be directly extrapolated to the nonlinear case. Let us review how the simplest knots emerge in this context. Consider, for instance, the pair
\begin{equation}
\alpha=\frac{r^{2}-t^2-1+2iz}{r^{2}-(t-i)^2}\quad\quad\beta=\frac{2(x-iy)}{r^{2}-(t-i)^2}
\end{equation}
which satisfies Eqs. (\ref{Bateman1}) and (\ref{Bateman2}). Here $(t,x,y,z)$ denotes usual cartesian coordinates in $\textbf{M}$ and $r^{2}=x^2+y^2+z^2$. In Ref. \cite{Ked} it is shown that the choice $f=\alpha^{p}$ and $g=\beta^{q}$ with $p,q\in\mathbb{Z}$ gives rise to fields whose electric and magnetic field lines are grouped into knotted and linked tori, nested one inside the other with $(p,q)$-torus knots at the core of the fibration. These being null fields satisfying Maxwell's equations, they automatically satisfy the quasilinear PDE's (\ref{NLE}). 

Also, the topological structure of the solutions is preserved for all times since, as shown by Irvine in \cite{Irv1}, the field lines satisfy the `frozen field' condition and evolve as if they were unbreakable filaments embedded in a fluid, stretching and deforming smoothly. In other words, the field as a whole evolve as an ambient isotopy of the spacelike hypersurface $\Sigma_{t}$. As an electromagnetic field has an infinite number of magnetic/electric field lines it is common to use two averaged quantities in order to quantify how much the field lines are knotted and linked. They are the \textit{electromagnetic helicities}, and are given explicitly in terms of the Whitehead integrals \cite{Whi}:
\begin{equation}\nonumber
\mathcal{H}_{m}:=\int {}^{\star}F^{ab}A_{a}t_{b}\ d^{3}x\quad\quad \mathcal{H}_{e}:=\int F^{ab}C_{b}t_{a}\ d^{3}x
\end{equation}
with $F_{ab}=\partial_{[a}A_{b]}$ and ${}^{\star}F_{ab}=\partial_{[a}C_{b]}$. These quantities are gauge invariant when the integral is over all space and one can show using the null condition that they do not depend on the choice of foliation. Therefore, the helicities are topological invariants and, for the above choice of $f$ and $g$, they read as $\mathcal{H}_{m}=\mathcal{H}_{e}=(p+q)^{-1}$ in appropriate units. The simplest solution $p=q=1$ is equivalent to a solution previously considered by Trautman and generalized by Bialynicki-Birula \cite{Ioa}. The lines of force associated to the $(p,q)$-solutions may be obtained by computing the integral curves of the \textit{Riemann-Silberstein vector}. These curves lie on the isosurfaces of the complex scalar \cite{Hoy}
\begin{equation}
\Phi_{E}+i\Phi_{B}=\alpha^{p}\beta^{q}
\end{equation}
and consist of knotted tori when $p\neq 1$, $q\neq 1$ are coprime integers. We note that the solutions hold as far as the nonlinear theory is well defined for null fields, i.e. $F=G=0$. Therefore, their universality do not depend on the specific choice of the Lagrangian.\\
 
\textbf{Probe waves and effective metrics} A direct consequence of nonlinear electrodynamics is the violation of the principle of superposition. This property implies that linearized wavy disturbances about a smooth background solution propagate nontrivially. Borrowing from the terminology of laser optics, we may say that \textit{probe fields} (waves) interact with \textit{pump fields} (knots) and are scattered by the latter. In this vein, a multitude of effects on the polarisation, wave-covector, frequency and velocity of `photons' that probe the background knots may take place. This is contrast with previous results due to Array\'as and Trueba, who investigated the classical relativistic motion of charged particles in a knotted electromagnetic field within the context of the linear theory \cite{Trueba}. They conclude that a deeper understanding of the interaction between electromagnetic knots and charged test particles could be useful to design experiments to produce knots in the laboratory. Interestingly, if the knots were produced in a nonlinear regime, we could naturally use incoming light from a distant source to probe the interaction.

Particularly, small amplitude/high-frequency perturbations of Eqs. (\ref{NLE}) are governed by two \textit{effective metrics} \cite{ngeo}. In this limit, the wave normals associated to the probe wave satisfy the modified dispersion relations (also called secular equations)
\begin{equation}\label{contrametrics}
\hat{g}_{\pm}^{ab}k_{a}k_{b}=0,
\end{equation}
with the reciprocal effective metric given by
\begin{equation}\label{pm}
\hat{g}_{\pm}^{ab}=g^{ab}+\mathcal{A}_{\pm}F^{ac}F_{c}^{\phantom a b}.
\end{equation}
Here $\mathcal{A}_{\pm}$ denotes a pair of functions (whose explicit form is unnecessary for our qualitative discussion) depending on the Lagrangian, background field and wave polarization. Thus, one may say that the background solutions behave as a nonlinear optical material with a response leading to birefringence. Birefringence is one of the fascinating properties of the vacuum of quantum electrodynamics which has been investigated with optical-cavity experiments.

Quite surprising, the effective metrics Eq. (\ref{pm}) associated to the electromagnetic knots may be recast in terms of the Robinson congruence determined by the null background. Due to Robinson's theorem \cite{Rob}, for any real, nonzero, null bivector $F_{kl}$, the conditions
\begin{equation}
F_{[kl}\sigma_{m]}=0,\quad\quad F_{km}\sigma^{m}=0.
\end{equation}
determine a geodesic and shear-free null congruence in spacetime. Conversely, if $\sigma^{m}$ is given, $F_{kl}$ can be determined up to a change of amplitude and polarization. It turns out that the reciprocal effective metric can be written as
\begin{equation}
\hat{g}_{\pm}^{ab}=g^{ab}+\mathcal{A}_{\pm}\sigma^{a}\sigma^{b}.
\end{equation}
Now, hyperbolicity guarantees that Eqs. (\ref{contrametrics}) describe algebraic varieties with the topologies of convex cones. We conclude from the last identity that these cones are locally determined by the vector field which generates the Robinson congruence. In terms of the electromagnetic fields, the latter read as
\begin{equation}
\sigma^{a}=\frac{1}{2}(E^{2}+B^{2})\left(t^{a}+\eta^{abcd}e_{b}b_{c}t_{d}\right).
\end{equation}
with $e_{b}$ and $b_{c}$ unit vectors in the directions of the electric and magnetic fields, respectively. 

With the co-cones in hand, we can readily predict the behaviour of incoming rays propagating on top of the background knot. They are given as the null geodesics of the \textit{effective metric}
\begin{equation}
\hat{g}_{ab}:=g_{ab}-\mathcal{A}_{\pm}\sigma_{a}\sigma_{b}, 
\end{equation}
which satisfies the relation $\hat{g}_{ac}\hat{g}^{cb}=\delta^{a}_{\phantom a b}$. This result suggests that the rays may receive an imprint of knottedness due to the interaction and, therefore, could be experimentally detected with future generations of super high power lasers such as the XFEL laser at DESY. Importantly,  as different kinds of knots cannot be deformed continuously into each other, the causal structure associated to each of them carries implicitly the topological invariant, which may be used to detect different knot signatures.\\

\textbf{Summary and prospects} The theoretical prediction of knotted states of light within the context of Maxell's theory and their possible experimental verification have received considerable attention in the last few years. In this paper we have shown that several types of known electromagnetic knots can be extended to nonlinear regimes in a quite simple way. Our result implies that probe waves arriving from a distant source would interact with the background knots nontrivially, at least if the latter were made of sufficiently intense fields. Given the fact that the behaviour of the rays are controlled by the Robinson congruence, it is fairly possible that distinct knots will imprint different degrees of knottedness to the incoming beam. A particularly exciting prospect is to investigate the behaviour of such rays in the context of the Euler-Heisenberg lagrangian, which takes into account one loop corrections due to QED. We hope that our results complement previous analysis and can be of interest in the experimental studies of light on top of intense fields.\\

\textbf{Acknowledgements} I would like to thank FAPESP grant 2011/11973-4 for funding my visit to ICTP-SAIFR from Month-Month 2015 where part of this work was done and CAPES for financial support.

\end{document}